\documentclass[aps,prd,onecolumn,showpacs,showkeys,amsmath,amssymb]{revtex4}
\usepackage{graphicx}
\usepackage{dcolumn}
\usepackage{bm}

\begin{document}
\title{The $Y(2175)$ State in the QCD Sum Rule}
%
\author{Hua-Xing Chen$^{1,2}$}
\email{hxchen@rcnp.osaka-u.ac.jp}
\author{Xiang Liu$^1$}
\email{xiangliu@pku.edu.cn}
\author{Atsushi Hosaka$^{2}$}
\email{hosaka@rcnp.osaka-u.ac.jp}
\author{Shi-Lin Zhu$^{1}$}
\email{zhusl@phy.pku.edu.cn}
\affiliation{$^1$Department of Physics, Peking University, Beijing
100871, China \\ $^2$Research Center for Nuclear Physics, Osaka
University, Ibaraki 567--0047, Japan}
\begin{abstract}
We study the mass of the state $Y(2175)$ of $J^{PC} = 1^{--}$ in the
QCD sum rule. We construct both the diquark-antidiquark currents
$(ss)(\bar s \bar s)$ and the meson-meson currents $(\bar ss)(\bar
ss)$. We find that there are two independent currents for both
cases, and derive the relations between them. The OPE convergence of
these two currents is sufficiently fast, which enables us to perform
good sum rule analysis. Both the SVZ sum rule and the finite energy
sum rule lead to a mass around $2.3 \pm 0.4$ GeV, which is
consistent with the observed mass within the uncertainties of the
present QCD sum rule. The coupling of the four-quark currents to
lower lying states such as $\phi(1020)$ turns out to be rather
small. We also discuss possible decay properties of $Y(2175)$ if it
is a tetraquark state.
\end{abstract}
\maketitle
\pagenumbering{arabic}
%
%
%
\section{Introduction}\label{sec:intro}
%

The theory of the strong interactions, Quantum Chromodynamics (QCD),
originated from the systematics of hadron spectroscopy. The
spectroscopy contains meson and baryon states, many of which are
well classified by the quark model with the quark content $q \bar q$
and $qqq$. Besides the quark model, QCD allows much richer hadron
spectrum such as multiquark states, hadron molecules, hybrid states,
glueballs etc. However the spectrum of QCD seems to saturate at $ q
\bar q$ and $qqq$. Since 2003, there have been important
developments in hadron spectroscopy, which is triggered by the
observation of the pentaquark $\Theta^+$. After three years of
intense study, the status of $\Theta^+$ is still controversial.
However, we have the charm-strange mesons $D_{sJ}(2317)$,
$D_{sJ}(2460)$ \cite{2317,2460}; the charmonium state
$X(3872)$~\cite{3872}, $Y(4260)$~\cite{4260}, and many $X$s and
$Y$s, whose properties seem difficult to be explained by the
conventional picture of $ q\bar q$.

Recently Babar Collaboration observed a resonance $Y(2175)$ near
the threshold in the process $e^+ e^- \rightarrow \phi f_0(980)$
via initial-state
radiation~\cite{Aubert:2006bu,Aubert:2007ur,Aubert:2007ym}. It has
the quantum numbers $J^{PC} = 1^{--}$. The Breit-Wigner mass is $M
= 2.175 \pm 0.010 \pm 0.015$ GeV, and width is $\Gamma = 0.058 \pm
0.016 \pm 0.020$ GeV. It has been also confirmed by BES
collaboration in the process $J / \psi \rightarrow \eta \phi
f_0(980)$. A fit with a Breit-Wigner function gives the peak mass
and width of $M=2.186\pm0.010\pm 0.006$ GeV and
$\Gamma=0.065\pm0.023\pm0.017$ GeV~\cite{:2007yt}.

There are many suggestions to interpret this resonance. Ding and Yan
interpreted it as a strangeonium hybrid and studied its decay
properties in the flux-tube model and the constituent gluon model.
Furthermore, for testing $s\bar{s}g$ scenario, they suggested
searching decay modes such as $Y(2175)\to K_1(1400)K\to \pi
K^{*}(892)K$, $Y(2175)\to K_1(1270)K\to \rho KK$ and $Y(2175)\to
K_{1}(1270)K\to \pi K_0^*(1430)$~\cite{Ding:2006ya}. In
Ref.~\cite{Ding-Yan}, the authors explored $Y(2175)$ as a $2^3D_1$
$s\bar{s}$ meson, and calculated its decay modes by using both the
$^{3}P_0$ model and the flux-tube model. They suggested experimental
search of the decay modes $KK$, $K^*K^*$, $K(1460)K$ and
$h_1(1380)\eta$. The characteristic decay modes of $Y(2175)$ as
either a hybrid state or an $s\bar{s}$ state are quite different,
which may be used to distinguish the hybrid and $s\bar{s}$ schemes.
Wang studied $Y(2175)$ as a tetraquark state $ss\bar s \bar s$ by
using QCD sum rule and suggested that there may be some tetraquark
components in the state $Y(2175)$~\cite{Wang:2006ri}. In a recent
article~\cite{Zhu:2007wz}, Zhu reviewed $Y(2175)$ and indicated that
the possibility of $Y(2175)$ arising from S-wave threshold effects
can not be excluded. Napsuciale, Oset, Sasaki and Vaquera-Araujo
studied the reaction $e^+e^-\to\phi\pi\pi$ for pions in an isoscalar
S-wave channel which is dominated by the loop mechanism. By
selecting the $\phi f_0(980)$ contribution as a function of the
$e^+e^-$ energy, they also reproduced the experimental data except
for the narrow peak~\cite{Napsuciale:2007wp}. Bystritskiy, Volkov,
Kuraev, Bartos and Secansky calculated the total probability and the
differential cross section of the process $e^+e^-\to \phi f_0(980)$
by using the local NJL model~\cite{Bystritskiy:2007wq}. Anikin, Pire
and Teryaev studied the reaction $\gamma^* \gamma \to \rho \rho$,
and calculated the mass of the isotensor exotic
meson~\cite{Anikin:2005ur}. In Ref.~\cite{Guo:2007uz}, the authors
performed a QCD sum rule study for $1^{--}$ hybrid meson, and the
mass is predicted to be $2.3-2.4$, $2.3-2.5$, and $2.5-2.6$ GeV for
$q \bar q g$, $q \bar s g$, and $s \bar s g$, respectively.

In this work, we revisit the possibility of $Y(2175)$ as an
tetraquark state $s s \bar s \bar s$. With the approach developed in
our previous work~\cite{Chen:2007zzg}, we construct the general
tetraquark interpolating currents with the quantum numbers $J^{PC} =
1^{--}$. We find that there are two independent currents. They can
have a structure of diquark-antidiquark $(ss)(\bar s \bar s)$, or
have a structure of meson-meson $(\bar s s)(\bar s s)$. We show that
they are equivalent, and derive the relations between them. Then by
using these two independent currents, we also perform a QCD sum rule
analysis. We calculate the OPE up to the dimension 12, which
contains the $\langle \bar q q \rangle^4$ condensates. In these two
respects, our study differs from the previous one of
Ref.~\cite{Wang:2006ri}.

This paper is organized as follows. In Sec.~\ref{sec:current}, we
construct the tetraquark currents using both diquark ($qq$) and
antidiquark ($\bar q \bar q$) fields, as well as quark-antiquark
($\bar q q$) pairs. In Sec.~\ref{sec:sumrule}, we perform a QCD
sum rule analysis by using these currents. In
Sec.~\ref{sec:numerical}, the numerical result is obtained for the
mass of $Y(2175)$. In Sec.~\ref{sec:fesr}, we use finite energy
sum rule to calculate its mass again. Sec.~\ref{sec:summary} is a
summary.

%
\section{Interpolating Currents}\label{sec:current}
%

In this section, we construct currents for the state $Y(2175)$ of
$J^{PC} = 1^{--}$. From the decay pattern $Y(2175) \to \phi(1020)
f_0(980)$, we expect that there is a large $ss \bar s \bar s$
component in $Y(2175)$. We may add further quark and antiquark
pairs, but the simplest choice would be $ss \bar s \bar s$. We will
discuss later how this simplest quark content will be compatible
with the above decay pattern when considering the possible structure
of $\phi(1020)$ and $f_0(980)$.

Let us now briefly see the flavor structure of the current. In the
diquark-antidiquark construction $(ss) (\bar s \bar s)$ where $ss$
and $\bar s \bar s$ pairs have a symmetric flavor structure, the
flavor decomposition goes as
\begin{equation}
\mathbf{6_f} \otimes
\mathbf{\bar 6_f} = \mathbf{1_f} \oplus \mathbf{8_f} \oplus
\mathbf{27_f} \, .
\end{equation}
Therefore, the $(ss) (\bar s \bar s)$ state is a mixing of $1_f,
8_f$ and $27_f$ multiplets in the ideal mixing scheme.

Now we find that there are two non-vanishing currents for each state
with the quantum number $J^{PC} = 1^{--}$. For the state $s s \bar s
\bar s$:
%
\begin{eqnarray}
\eta_{1\mu} &=& (s_a^T C \gamma_5 s_b) (\bar{s}_a \gamma_\mu
\gamma_5 C \bar{s}_b^T) - (s_a^T C \gamma_\mu \gamma_5 s_b)
(\bar{s}_a \gamma_5 C \bar{s}_b^T) \label{def:eta1} \, ,
\\ \eta_{2\mu} &=& (s_a^T C \gamma^\nu s_b) (\bar{s}_a \sigma_{\mu\nu} C \bar{s}_b^T)
- (s_a^T C \sigma_{\mu\nu} s_b) (\bar{s}_a \gamma^\nu C \bar{s}_b^T)
\label{def:eta2} \, ,
\end{eqnarray}
%
where the sum over repeated indices ($\mu$ for Dirac spinor
indices, and $a, b$ for color indices) is taken. $C = i\gamma_2
\gamma_0$ is the Dirac field charge conjugation operator, and the
superscript $T$ represents the transpose of the Dirac indices
only.

Besides the diquark-antidiquark currents, we can also construct the
tetraquark currents by using quark-antiquark ($\bar s s$) pairs. We
find that there are four non-vanishing currents:
%
\begin{eqnarray} \nonumber
\eta_{3\mu} &=& (\bar{s}_a s_a)(\bar{s}_b \gamma_\mu s_b) \, ,
\\ \nonumber \eta_{4\mu} &=& (\bar{s}_a \gamma^\nu\gamma_5 s_a)(\bar{s}_b \sigma_{\mu\nu}\gamma_5
s_b) \, ,
\\ \nonumber \eta_{5\mu} &=& {\lambda_{ab}}{\lambda_{cd}}(\bar{s}_a s_b)(\bar{s}_c
\gamma_\mu s_d) \, ,
\\ \nonumber \eta_{6\mu} &=& {\lambda_{ab}}{\lambda_{cd}}
(\bar{s}_a \gamma^\nu\gamma_5 s_b)(\bar{s}_c \sigma_{\mu\nu}\gamma_5
s_d) \, .
\end{eqnarray}
%
In Ref.~\cite{Wang:2006ri}, the author used $\eta_{5\mu}$ to perform
QCD sum rule analysis, which is a mixing of $\eta_{1\mu}$ and
$\eta_{2\mu}$. We can verify the following relations by using the
Fierz transformation:
%
\begin{eqnarray}
\eta_{5\mu} = - \frac{5}{3} \eta_{3\mu} - i \eta_{4\mu} \, , \, \,
\eta_{6\mu} = 3i \eta_{3\mu} + \frac{1}{3} \eta_{4\mu} \, .
\end{eqnarray}
%
Therefore, among the four $(\bar q q)(\bar q q)$ currents, two are
independent. We can also verify the relations between $(ss)(\bar
s\bar s)$ currents and $(\bar s s)(\bar s s)$ currents, by using the
Fierz transformation:
%
\begin{eqnarray}
\eta_{1\mu} = - \eta_{3\mu} + i \eta_{4\mu} \, , \, \, \eta_{2\mu} =
3i \eta_{3\mu} - \eta_{4\mu} \, .
\end{eqnarray}
%
Therefore, these two constructions are equivalent, and we will use
$\eta_{1\mu}$ and $\eta_{2\mu}$ for QCD sum rule analysis.

%
\section{QCD sum rule Analysis}\label{sec:sumrule}
%

For the past decades QCD sum rule has proven to be a very powerful
and successful non-perturbative
method~\cite{Shifman:1978bx,Reinders:1984sr}, and it has been
applied to study tetraquark states in many
references~\cite{Chen:2007zzg,Zhang:2006xp,Matheus:2006xi,Lee:2007mva,Sugiyama:2007sg,Matheus:2007ta,Lee:2007gs}.
In sum rule analyses, we consider two-point correlation functions:
%
\begin{equation}
\Pi_{\mu\nu}(q^2) \, \equiv \, i \int d^4x e^{iqx} \langle 0 | T
\eta_\mu(x) { \eta_\nu^\dagger } (0) | 0 \rangle \, , \label{def:pi}
\end{equation}
%
where $\eta_\mu$ is an interpolating current for the tetraquark. The
Lorentz structure can be simplified to be:
%
\begin{equation}
\Pi_{\mu\nu}(q^2) = ( {q_\mu q_\nu \over q^2} - g_{\mu\nu} )
\Pi^{(1)}(q^2) + {q_\mu q_\nu \over q^2} \Pi^{(0)}(q^2) \, .
\label{def:pi1}
\end{equation}
%

We compute $\Pi(q^2)$ in the operator product expansion (OPE) of QCD
up to certain order in the expansion, which is then matched with a
hadronic parametrization to extract information of hadron
properties. At the hadron level, we express the correlation function
in the form of the dispersion relation with a spectral function:
%
\begin{equation}
\Pi^{(1)}(q^2)=\int^\infty_{16
m_s^2}\frac{\rho(s)}{s-q^2-i\varepsilon}ds \, , \label{eq:disper}
\end{equation}
%
where the subscript is $(4 m_s)^2 = 16 m_s^2$, and
%
\begin{eqnarray}
\rho(s) & \equiv & \sum_n\delta(s-M^2_n)\langle
0|\eta|n\rangle\langle n|{\eta^\dagger}|0\rangle \ \nonumber\\ &=&
f^2_Y\delta(s-M^2_Y)+ \rm{higher\,\,states}\, . \label{eq:rho}
\end{eqnarray}
%
For the second equation, as usual, we adopt a parametrization of one
pole dominance for the ground state $Y$ and a continuum
contribution. The sum rule analysis is then performed after the
Borel transformation of the two expressions of the correlation
function, (\ref{def:pi}) and (\ref{eq:disper})
%
\begin{equation}
\Pi^{(all)}(M_B^2)\equiv\mathcal{B}_{M_B^2}\Pi^{(1)}(p^2)=\int^\infty_{16
m_s^2} e^{-s/M_B^2} \rho(s)ds \, . \label{eq_borel}
\end{equation}
%
Assuming the contribution from the continuum states can be
approximated well by the spectral density of OPE above a threshold
value $s_0$ (duality), we arrive at the sum rule equation
%
\begin{equation}
\Pi(M_B^2) \equiv f^2_Y e^{-M_Y^2/M_B^2} = \int^{s_0}_{16 m_s^2}
e^{-s/M_B^2}\rho(s)ds \label{eq_fin} \, .
\end{equation}
%
Differentiating Eq.~(\ref{eq_fin}) with respect to $1 / M_B^2$ and
dividing it by Eq. (\ref{eq_fin}), finally we obtain
%
\begin{equation}
M^2_Y =
\frac{\frac{\partial}{\partial(-1/M_B^2)}\Pi(M_B^2)}{\Pi(M_B^2)} =
\frac{\int^{s_0}_{16 m_s^2} e^{-s/M_B^2}s\rho(s)ds}{\int^{s_0}_{16
m_s^2} e^{-s/M_B^2}\rho(s)ds}\, . \label{eq_LSR}
\end{equation}
%
In the following, we study both Eqs.~(\ref{eq_fin}) and
(\ref{eq_LSR}) as functions of the parameters such as the Borel mass
$M_B$ and the threshold value $s_0$ for various combinations of the
tetraquark currents.

For the currents $\eta_{1\mu}$ and $\eta_{2\mu}$, we have calculated
the OPE up to dimension twelve, which contains the $\langle \bar q q
\rangle^4$ condensate:
%
\begin{eqnarray}
\Pi_1(M_B^2) &=& \int^{s_0}_{16 m_s^2} \Bigg [ {s^4 \over 18432
\pi^6} - { m_s^2 s^3 \over 256 \pi^6 } + \Big ( - { \langle g^2 G
G \rangle \over 18432 \pi^6 } + {m_s \langle \bar s s \rangle
\over 48 \pi^4} \Big ) s^2 \nonumber\\  && + \Big ( { \langle \bar
s s \rangle^2 \over 18 \pi^2 } - { m_s \langle g \bar s \sigma G s
\rangle \over 48 \pi^4 } + { 17 m_s^2 \langle g^2 G G \rangle
\over 9216 \pi^6 } \Big ) s + \Big ( { \langle \bar s s \rangle
\langle g \bar s \sigma G s \rangle \over 12 \pi^2 } - { m_s
\langle g^2 G G \rangle \langle \bar s s \rangle \over 128 \pi^4}
- { 29 m_s^2 \langle \bar s s
\rangle^2 \over 12 \pi^2 } \Big ) \Bigg ] e^{-s/M_B^2} ds \nonumber\\
 && + \Big (  {5 \langle g^2 GG \rangle \langle \bar s s
\rangle^2 \over 864 \pi^2} + { \langle g \bar s \sigma G s \rangle^2
\over 48 \pi^2 } + {20 m_s \langle \bar s s \rangle^3 \over 9} - {5
m_s \langle g^2 GG \rangle \langle g \bar s \sigma G s \rangle \over
2304 \pi^4 } - { 3 m_s^2 \langle \bar s s \rangle
\langle g \bar s \sigma G s \rangle \over 2 \pi^2 } \Big ) \nonumber\\
 && + {1 \over M_B^2} \Big ( - { 32 g^2 \langle \bar s s
\rangle^4 \over 81 } - { \langle g^2 GG \rangle \langle \bar s s
\rangle \langle g \bar s \sigma G s \rangle \over 576 \pi^2 } - {
10 m_s \langle \bar s s \rangle^2 \langle g \bar s \sigma G s
\rangle \over 9 } + { m_s^2 \langle g^2 GG \rangle \langle \bar s
s \rangle^2 \over 576 \pi^2 } + { m_s^2 \langle g \bar s \sigma G
s \rangle^2 \over 12 \pi^2 } \Big ) \label{eq:pimb1} \, ,\nonumber\\
\\ \Pi_2(M_B^2) &=& \int^{s_0}_{16 m_s^2} \Bigg [ {s^4 \over 12288 \pi^6}
- { 3 m_s^2 s^3 \over 512 \pi^6 } + \Big ( { \langle g^2 G G \rangle
\over 18432 \pi^6 } + {m_s \langle \bar s s \rangle \over 32 \pi^4} \Big ) s^2 \nonumber\\
 && + \Big ( { \langle \bar s s \rangle^2 \over 12 \pi^2 }
- { m_s \langle g \bar s \sigma G s \rangle \over 32 \pi^4 } + { 35
m_s^2 \langle g^2 G G \rangle \over 9216 \pi^6 } \Big ) s + \Big ( {
\langle \bar s s \rangle \langle g \bar s \sigma G s \rangle \over 8
\pi^2 } - { 3 m_s \langle g^2 G G \rangle \langle \bar s s \rangle
\over 128 \pi^4} - { 29 m_s^2 \langle \bar s s
\rangle^2 \over 8 \pi^2 }  \Big ) \Bigg ] e^{-s/M_B^2} ds \nonumber\\
 && + \Big ( {5 \langle g^2 GG \rangle \langle \bar s s
\rangle^2 \over 288 \pi^2} + { \langle g \bar s \sigma G s
\rangle^2 \over 32 \pi^2 } + {10 m_s \langle \bar s s \rangle^3
\over 3} - {5 m_s \langle g^2 GG \rangle \langle g \bar s \sigma G
s \rangle \over 768 \pi^4 } - { 9 m_s^2 \langle \bar s s \rangle
\langle g \bar s \sigma G s \rangle \over 4 \pi^2 } \Big )
\nonumber\\  && + {1 \over M_B^2} \Big ( - { 16 g^2 \langle \bar s
s \rangle^4 \over 27 } - { \langle g^2 GG \rangle \langle \bar s s
\rangle \langle g \bar s \sigma G s \rangle \over 192 \pi^2 } - {
5 m_s \langle \bar s s \rangle^2 \langle g \bar s \sigma G s
\rangle \over 3 } - { m_s^2 \langle g^2 GG \rangle \langle \bar s
s \rangle^2 \over 576 \pi^2 } + { m_s^2 \langle g \bar s \sigma G
s \rangle^2 \over 8 \pi^2 } \Big ) \label{eq:pimb2} \, .\nonumber\\
\end{eqnarray}
In the above equations, $\langle \bar{s}s \rangle$ is the dimension
$D=3$ strange quark condensate; $\langle g^2 GG \rangle$ is a $D=4$
gluon condensate; $\langle g\bar{s}\sigma Gs \rangle$ is $D=5$ mixed
condensate. There are many terms which give minor contributions,
such as $\langle g^3 G^3 \rangle$, and we omit them. As usual, we
assume the vacuum saturation for higher dimensional condensates such
as $\langle 0 | \bar q q \bar q q |0\rangle \sim \langle 0 | \bar q
q |0\rangle \langle 0|\bar q q |0\rangle$. To obtain these results,
we keep the terms of order $O(m_q^2)$ in the propagators of a
massive quark in the presence of quark and gluon condensates:
%
\begin{eqnarray} \nonumber
i S^{ab} & \equiv & \langle 0 | T [ q^a(x) q^b(0) ] | 0 \rangle
\\ \nonumber &=& { i \delta^{ab} \over 2 \pi^2 x^4 } \hat{x} + {i \over
32\pi^2} { \lambda^n_{ab} \over 2 } g_c G^n_{\mu\nu} {1 \over x^2}
(\sigma^{\mu\nu} \hat{x} + \hat{x} \sigma^{\mu\nu}) - { \delta^{ab}
\over 12 } \langle \bar q q \rangle \\ && + { \delta^{ab} x^2 \over
192 } \langle g_c \bar q \sigma G q \rangle - { m_q \delta^{ab}
\over 4 \pi^2 x^2 } + { i \delta^{ab} m_q \langle \bar q q \rangle
\over 48 }  \hat x + { i \delta^{ab} m_q^2 \over 8 \pi^2 x^2 }
\hat{x}
\end{eqnarray}
%

We find that there is an approximate relation between the
correlation functions of $\eta_{1\mu}$ and $\eta_{2\mu}$:
\begin{equation}
3 \Pi_1(M_B^2) \sim 2 \Pi_2(M_B^2) \, , \label{eq:pi12}
\end{equation}
which is valid for the continuum, $\langle \bar s s \rangle$, and
$\langle g_c \bar q \sigma G q \rangle$ terms, etc. So the numerical
results by using them are also very similar.

%
\section{Numerical Analysis}\label{sec:numerical}
%

In our numerical analysis, we use the following values for various
condensates and $m_s$ at 1 GeV and $\alpha_s$ at 1.7 GeV
~\cite{Yang:1993bp,Narison:2002pw,Gimenez:2005nt,Jamin:2002ev,Ioffe:2002be,Ovchinnikov:1988gk,Ellis:1996xc,Yao:2006px}:
%
\begin{eqnarray}
\nonumber &&\langle\bar qq \rangle=-(0.240 \mbox{ GeV})^3\, ,
\\
\nonumber &&\langle\bar ss\rangle=-(0.8\pm 0.1)\times(0.240 \mbox{
GeV})^3\, ,
\\
\nonumber &&\langle g_s^2GG\rangle =(0.48\pm 0.14) \mbox{ GeV}^4\, ,
\\
\label{condensates} && \langle g_s\bar q\sigma G
q\rangle=-M_0^2\times\langle\bar qq\rangle\, ,
\\
\nonumber && M_0^2=(0.8\pm0.2)\mbox{ GeV}^2\, ,
\\
\nonumber &&m_s(1\mbox{ GeV})=125 \pm 20 \mbox{ MeV}\, ,
\\
\nonumber && \alpha_s(1.7\mbox{GeV}) = 0.328 \pm 0.03
\pm 0.025 \, .
\end{eqnarray}
%
There is a minus sign in the definition of the mixed condensate
$\langle g_s\bar q\sigma G q\rangle$, which is different from that
used in some other QCD sum rule studies. This difference just comes
from the definition of coupling constant
$g_s$~\cite{Yang:1993bp,Hwang:1994vp}.

\begin{figure}[hbt]
\begin{center}
\scalebox{0.85}{\includegraphics{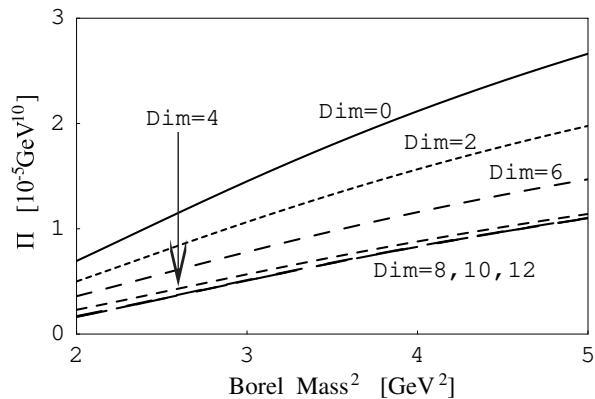}} \caption{Various
contribution to the correlation function for the current
$\eta_{1\mu}$ as functions of the Borel mass $M_B$ in units of
GeV$^{10}$ at $s_0 = 5.7$ GeV$^2$. The labels indicate the dimension
up to which the OPE terms are included.} \label{fig:convergence}
\end{center}
\end{figure}

First we want to study the convergence of the operator product
expansion, which is the cornerstone of the reliable QCD sum rule
analysis. By taking $s_0$ to be $\infty$ and the integral
subscript $16 m_s^2$ to be zero, we obtain the numerical series of
the OPE as a function of $M_B$:
\begin{eqnarray} \nonumber
\Pi_1(M_B^2) &=& 1.4 \times 10^{-6} M_B^{10} - 3.8 \times 10^{-7}
M_B^8 - 6.2 \times 10^{-7} M_B^6 + 4.2 \times 10^{-7} M_B^4 \\
&& - 1.2 \times 10^{-6} M_B^2 + 4.7 \times 10^{-8} - 1.5 \times
10^{-7} M_B^{-2} \, ,
\\ \nonumber
\Pi_2(M_B^2) &=& 2.0 \times 10^{-6} M_B^{10} - 5.7 \times 10^{-7}
M_B^8 - 8.0 \times 10^{-7} M_B^6 + 6.4 \times 10^{-7} M_B^4 \\
&& - 1.7 \times 10^{-6} M_B^2 + 1.0 \times 10^{-7} - 2.2 \times
10^{-7} M_B^{-2} \, .
\end{eqnarray}
After careful testing of the free parameter Borel mass $M_B$, we
find for $M_B^2 > 2$ GeV$^2$, which is the region suitable for the
study of $Y(2175)$, the Borel mass dependence is weak. Moreover, the
convergence of the OPE is satisfied in this region. The correlation
function of the current $\eta_{1\mu}$ is shown in
Fig.~\ref{fig:convergence}, when we take $s_0 = 5.7$ GeV$^2$ (the
integral subscript is still $16 m_s^2$). We find that in the region
of 2 GeV$^2 < M_B^2 < 5$ GeV$^2$, the perturbative term (the solid
line in Fig.~\ref{fig:convergence}) gives the most important
contribution, and the convergence is quite good.
%
%

It is important to note that the $Y(2175)$ state is not the lowest
state in the $1^{--}$ channel containing $s\bar s$ and that the
interpolating currents see only the quantum number of the states. It
is possible that the low-lying states $\phi(1020)$ and $\phi(1680)$
also couple to the tetraquark currents $\eta_{1\mu}$ and
$\eta_{2\mu}$. If so, their contribution to the spectral density and
the resulting correlation function should be positive definite.

\begin{figure}[hbt]
\begin{center}
\scalebox{0.85}{\includegraphics{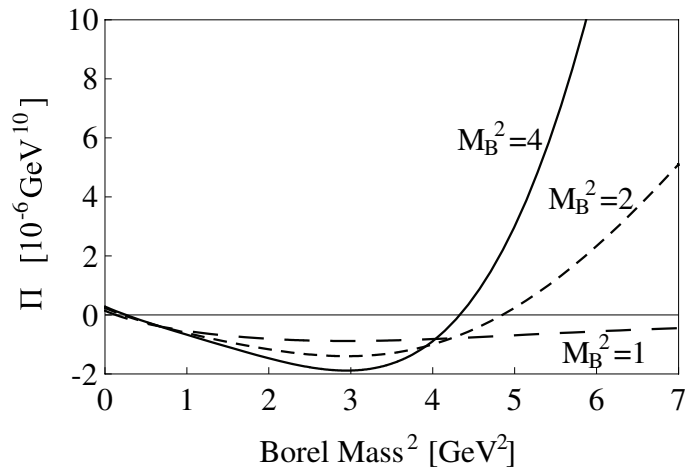}} \caption{The correlation
function for the current $\eta_{1\mu}$ as a function of $s_0$ in
units of GeV$^{10}$. The curves are obtained by setting $M_B^2 = 1$
GeV$^2$ (long-dashed line), 2 GeV$^2$ (short-dashed line) and 4
GeV$^2$ (solid line).} \label{fig:pi}
\end{center}
\end{figure}

However, we find that (1) the spectral densities $\rho(s)$ of
Eq.~(\ref{eq:rho}) for both currents $\eta_{1\mu}$ and $\eta_{2\mu}$
are negative when $s < 2$ GeV$^2$; (2) the Borel transformed
correlation function $\Pi(M_B^2)$ in Eq.~(\ref{eq_fin}) is also
negative in the region $s_0 < 4.3$ GeV$^2$ and $1$ GeV$^2 < M_B^2 <
4$ GeV$^2$. As an illustration, we show the correlation function as
a function of $s_0$ in Fig.~\ref{fig:pi}. This fact indicates that
the $s s \bar s \bar s$ tetraquark currents couple weakly to the
lower states $\phi(1020)$ and $\phi(1680)$ in the present QCD sum
rule analysis.
%
%

The pole contribution is not large enough for both currents due to
$D=10$ perturbative term $\int_0^{s_0} e^{-s/M_B^2} s^4 ds$, which
is a common feature for any multiquark interpolating currents with
high dimensions. The mixing of the currents $\eta_{1\mu}$ and
$\eta_{2\mu}$ does not improve the rate of the pole contribution.
The small pole contribution suggests that the continuum contribution
to the spectral density is dominant, which demands a very careful
choice of the parameters of the QCD sum rule. In our numerical
analysis, we require the extracted mass have a dual minimum
dependence on both the Borel parameter $M_B$ and the threshold
parameter $s_0$. In this way, we can find a good working region of
$M_B$ and $s_0$ (Borel window), where the mass of $Y(2175)$ can be
determined reliably.

\begin{figure}[hbt]
\begin{center}
\scalebox{0.85}{\includegraphics{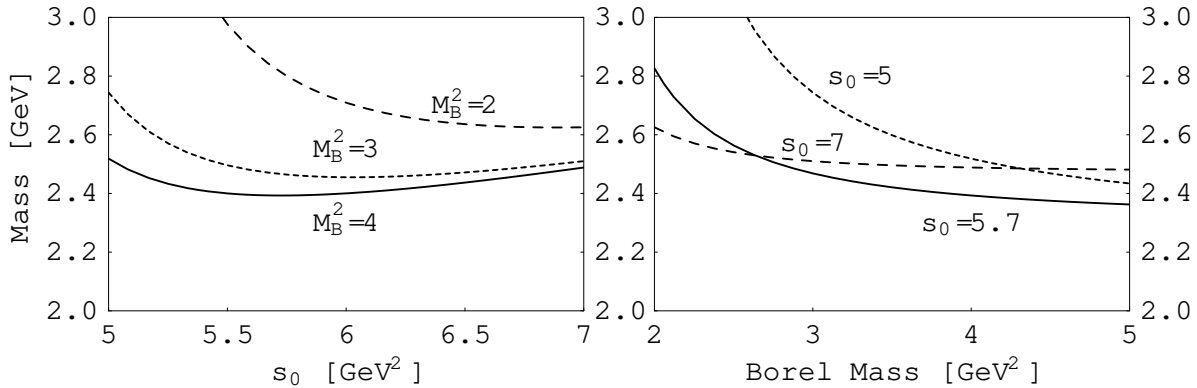}} \caption{The mass of
$Y(2175)$ as a function of $M_B$ (Left) and $s_0$ (Right) in units
of GeV for the current $\eta_{1\mu}$.} \label{fig:eta1}
\end{center}
\end{figure}

\begin{figure}[hbt]
\begin{center}
\scalebox{0.85}{\includegraphics{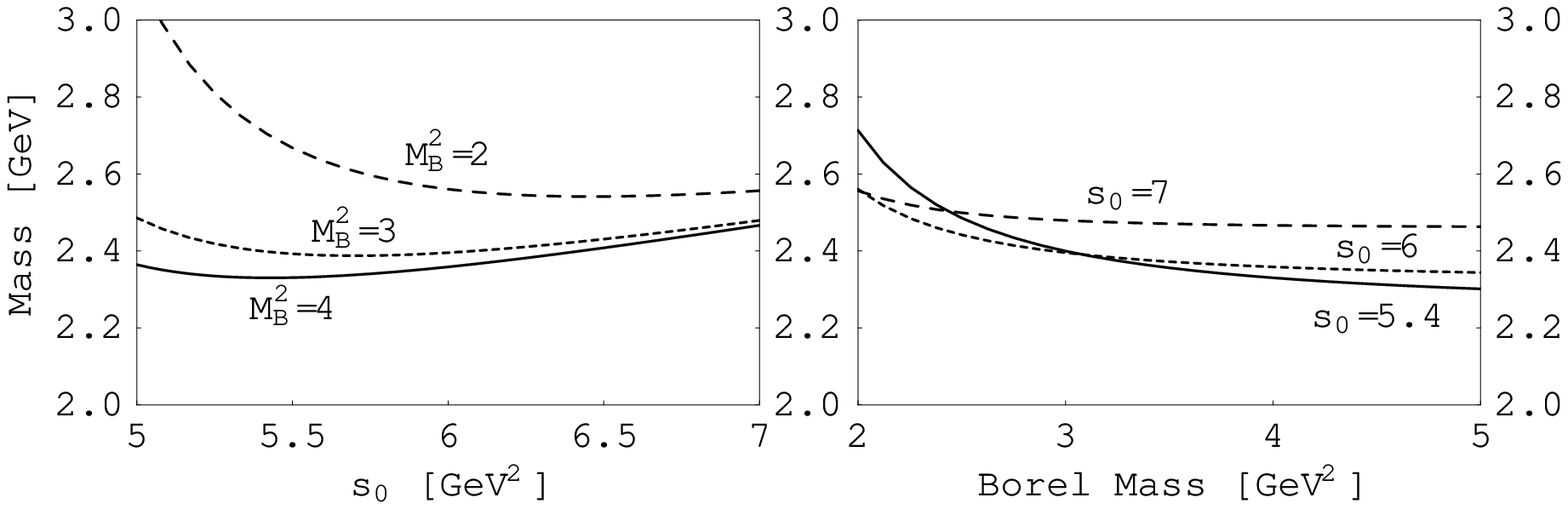}} \caption{The mass of
$Y(2175)$ as a function of $M_B$ (Left) and $s_0$ (Right) in units
of GeV for $\eta_{2\mu}$.} \label{fig:eta2}
\end{center}
\end{figure}

Now the mass is shown as functions of the Borel mass $M_B$ and the
threshold value $s_0$ in Fig.~\ref{fig:eta1} and
Fig.~\ref{fig:eta2}. The threshold value is taken to be around $5
\sim 7$ GeV$^2$, where its square root is around $2.2 \sim 2.7$ GeV.
We find that there is a mass minimum around 2.4 GeV for the current
$\eta_{1\mu}$, when we take $M_B^2 \sim 4$ GeV$^2$ and $s_0 \sim
5.7$ GeV$^2$. While this minimum is around 2.3 GeV for the current
$\eta_{2\mu}$, when we take $M_B^2 \sim 4$ GeV$^2$ and $s_0 \sim
5.4$ GeV$^2$.
%
%
%
%

In short summary, we have performed the QCD sum rule analysis for
both $\eta_{1\mu}$ and $\eta_{2\mu}$. The obtained results are
quite similar. This is due to the similarity of the two
correlation functions as shown in Eq.~(\ref{eq:pi12}). We have
also considered their mixing, which also give the similar result.
The mass is predicted to be around $2.3 \sim 2.4$ GeV in the QCD
sum rule.

%
\section{Finite Energy Sum Rule}\label{sec:fesr}
%

In this section, we use the method of finite energy sum rule
(FESR). In order to calculate the mass in the FESR, we first
define the $n$th moment by using the spectral function $\rho(s)$
in Eq.~(\ref{eq:rho})
%
\begin{equation}
W(n, s_0) = \int^{s_0}_0 \rho(s) s^n ds \, . \label{eq:moment}
\end{equation}
%
This integral is used for the phenomenological side, while the
integral along the circular contour of radius $s_0$ on the $q^2$
complex plain should be performed for the theoretical side. The
lower integral bound $s=0$ is taken in order to include the
$delta$-function contribution in the OPE (Eqs.~(\ref{eq:rho1}) and
(\ref{eq:rho2})).

With the assumption of quark-hadron duality, we obtain
%
\begin{equation}
W(n, s_0)\Big |_{Hadron} = W(n, s_0)\Big |_{OPE} \, .
\end{equation}
%
The mass of the ground state can be obtained as
%
\begin{equation}
M^2_Y(n, s_0)= { W(n+1, s_0) \over W(n, s_0) } \, . \label{eq_FESR}
\end{equation}
%
For the currents $\eta_{1\mu}$ and $\eta_{2\mu}$, the spectral
functions $\rho_1(s)$ and $\rho_2(s)$ can be drawn from
Eqs.~(\ref{eq:pimb1}) and (\ref{eq:pimb2}). The d = 12 terms which
are proportional to $1 / (q^2)^2$ do not contribute to the function
$W(n,s_0)$ of Eq.~(\ref{eq:moment}) for $n=0$, or they have a very
small contribution for $n=1$, when the theoretical side is computed
by the integral over the circle of radius $s_0$ on the complex $q^2$
plain. Therefore, the spectral densities for $\eta_{1\mu}$ and
$\eta_{2\mu}$ take the following form up to dimension 10,
%
\begin{eqnarray}
\rho_1(s) &=& {s^4 \over 18432 \pi^6} - { m_s^2 s^3 \over 256 \pi^6
} + \Big ( - { \langle g^2 G G \rangle \over 18432 \pi^6 } + {m_s
\langle \bar s s \rangle \over 48 \pi^4} \Big ) s^2 \nonumber\\  &&
+ \Big ( { \langle \bar s s \rangle^2 \over 18 \pi^2 } - { m_s
\langle g \bar s \sigma G s \rangle \over 48 \pi^4 } + { 17 m_s^2
\langle g^2 G G \rangle \over 9216 \pi^6 } \Big ) s + \Big ( {
\langle \bar s s \rangle \langle g \bar s \sigma G s \rangle \over
12 \pi^2 } - { m_s \langle g^2 G G \rangle \langle \bar s s \rangle
\over 128 \pi^4} - { 29 m_s^2 \langle \bar s s \rangle^2 \over 12
\pi^2 } \Big ) \nonumber \\ && + \Big (  {5 \langle g^2 GG \rangle
\langle \bar s s \rangle^2 \over 864 \pi^2} + { \langle g \bar s
\sigma G s \rangle^2 \over 48 \pi^2 } + {20 m_s \langle \bar s s
\rangle^3 \over 9} - {5 m_s \langle g^2 GG \rangle \langle g \bar s
\sigma G s \rangle \over 2304 \pi^4 } - { 3 m_s^2 \langle \bar s s
\rangle
\langle g \bar s \sigma G s \rangle \over 2 \pi^2 } \Big ) \delta(s) \label{eq:rho1} \, ,\nonumber\\
\\ \rho_2(s) &=& {s^4 \over 12288 \pi^6}
- { 3 m_s^2 s^3 \over 512 \pi^6 } + \Big ( { \langle g^2 G G \rangle
\over 18432 \pi^6 } + {m_s \langle \bar s s \rangle \over 32 \pi^4} \Big ) s^2 \nonumber\\
 && + \Big ( { \langle \bar s s \rangle^2 \over 12 \pi^2 }
- { m_s \langle g \bar s \sigma G s \rangle \over 32 \pi^4 } + { 35
m_s^2 \langle g^2 G G \rangle \over 9216 \pi^6 } \Big ) s + \Big ( {
\langle \bar s s \rangle \langle g \bar s \sigma G s \rangle \over 8
\pi^2 } - { 3 m_s \langle g^2 G G \rangle \langle \bar s s \rangle
\over 128 \pi^4} - { 29 m_s^2 \langle \bar s s
\rangle^2 \over 8 \pi^2 }  \Big ) \nonumber\\
&& + \Big ( {5 \langle g^2 GG \rangle \langle \bar s s \rangle^2
\over 288 \pi^2} + { \langle g \bar s \sigma G s \rangle^2 \over 32
\pi^2 } + {10 m_s \langle \bar s s \rangle^3 \over 3} - {5 m_s
\langle g^2 GG \rangle \langle g \bar s \sigma G s \rangle \over 768
\pi^4 } - { 9 m_s^2 \langle \bar s s \rangle
\langle g \bar s \sigma G s \rangle \over 4 \pi^2 } \Big ) \delta(s) \label{eq:rho2} \, .\nonumber\\
\end{eqnarray}
%

%
\begin{figure}[hbt]
\begin{center}
\scalebox{0.85}{\includegraphics{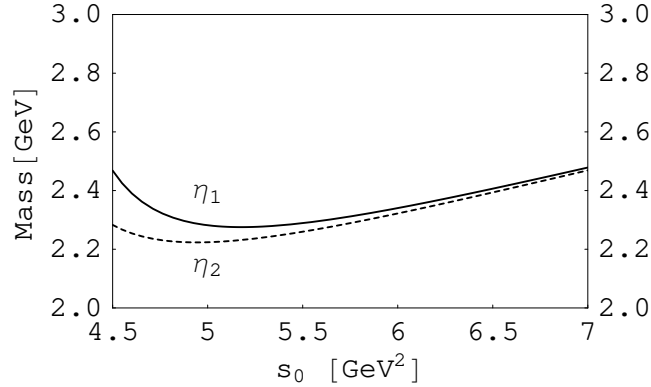}} \caption{The mass of
$Y(2175)$ by using the current $\eta_{1\mu}$ (solid line) and
$\eta_{2\mu}$ (dashed line) as a function of $s_0$ in units of GeV.}
\label{fig:fesr}
\end{center}
\end{figure}
%

\begin{figure}[hbt]
\begin{center}
\scalebox{0.85}{\includegraphics{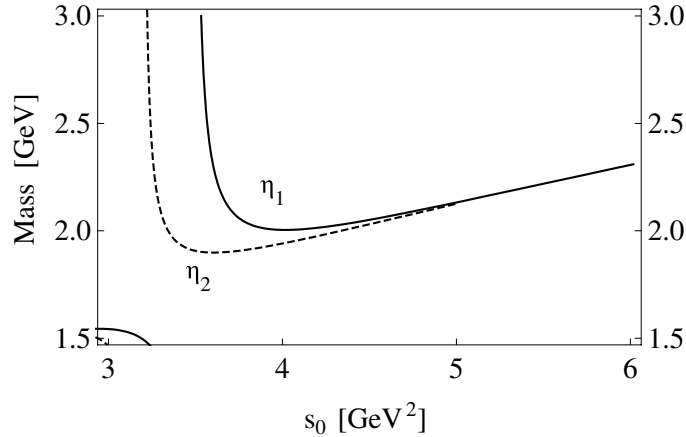}} \caption{The
mass of $Y(2175)$ by using the current $\eta_{1\mu}$ (solid line)
and $\eta_{2\mu}$ (dashed line) as a function of $s_0$ in units of
GeV, when the lower limit of integrations in Eq.~(\ref{eq_FESR}) is
2 GeV$^2$ instead of $16m_s^2$.} \label{fig:fesr_correct}
\end{center}
\end{figure}

The mass is shown as a function of the threshold value $s_0$ in
Fig.~\ref{fig:fesr}, where $n$ is chosen to be 1. We find that there
is a mass minimum. It is around 2.3 GeV for the current
$\eta_{1\mu}$ when we take $s_0 \sim 5.2$ GeV$^2$, while it is
around 2.2 GeV for the current $\eta_{2\mu}$ when we take $s_0 \sim
4.8$ GeV$^2$. For the current $\eta_{1\mu}$, the minimum point
occurs at $\sqrt{s_0} = 2.28$ GeV where the mass takes 2.3 GeV, and
the threshold value is slightly smaller than the mass, unlike the
ordinary expectation that $\sqrt{s_0}$ is larger than the obtained
mass. However, the minimum point is on the very shallow minimum
curve and the resulting mass is rather insensitive to the change in
the $\sqrt{s_0}$ value. Therefore, we can increase $\sqrt{s_0}$
slightly more, for example 2.45 GeV, but the mass still remains at
around $2.35$ GeV, which is smaller than $\sqrt{s_0}$ now. This fact
is due to the uncertainty of our sum rule analysis as well as the
negative part of the spectral densities. For example, if we take the
lower limit of integrations in Eq.~(\ref{eq_FESR}) to be 1 GeV$^2$
instead of $16m_s^2$ (the positive part starts at around 3 GeV$^2$),
the mass minimum will be around 2.1 GeV, when $s_0$ is around 4.5
GeV$^2$; if we take the lower limit to be 2 GeV$^2$, the mass
minimum would be around 2.0 GeV, when $s_0$ is around 4 GeV$^2$. We
show the second case in Fig.~\ref{fig:fesr_correct}. The region $5 <
s_0 < 6$ GeV$^2$ is suitable for the QCD sum rule analysis, and the
mass obtained is around 2.2 GeV. Therefore, considering the
uncertainty of the QCD sum rule, we obtain the same result as the
previous one.

%
%

%
\section{Summary and Discussions}\label{sec:summary}
%

In this work we have studied the mass of the state $Y(2175)$ with
the quantum numbers $J^{PC} = 1^{--}$ in the QCD sum rule. We have
constructed both the diquark-antidiquark currents $(ss)(\bar s \bar
s)$ and the meson-meson currents $(\bar s s)(\bar s s)$. We find
that there are two independent currents for both cases and verify
the relations between them. Then using the two $(ss)(\bar s \bar s)$
currents, we calculate the OPE up to dimension twelve, which
contains the $\langle \bar s s \rangle^4$ condensates. The
convergence of the OPE turns our to be very good. We find that the
OPE's of the two currents are similar, and therefore, the obtained
results are also similar. By using both the SVZ sum rule and the
finite energy sum rule, we find that there is a mass minimum. For
SVZ sum rule, the minimum is in the region $5 < s_0 < 7$ GeV$^2$ and
$2< M_B^2 < 4$ GeV$^2$. For finite energy sum rule, the minimum is
in the region $4.5 < s_0 < 5.5$ GeV$^2$. Considering the
uncertainty, the mass obtained is around $2.3\pm0.4$ GeV. The state
$Y(2175)$ can be accommodated in the QCD sum rule formalism although
the central value of the mass is about 100 MeV higher than the
experimental value. We calculate the OPE up to dimension twelve and
include many terms, but still the accuracy is around 20\%. This is
the usual accuracy of the QCD sum rule. In our analysis it is partly
due to the many omitted condensates such as $\langle GGG \rangle$
etc.

We have investigated the coupling of the currents to the lower lying
states including $\phi(1020)$ and found that the relevant spectral
density becomes negative, implying that the present four-quark
currents can not describe those states properly. This fact indicates
that the four-quark interpolating currents couple rather weakly to
$\phi(1020)$, which is a pure $s\bar s$ state.

We can test the tetraquark structure of $Y(2175)$ by considering its
decay properties. Naively, the $ss \bar s \bar s$ tetraquark would
fall apart via $S$-wave into the $\phi(1020)f_0(980)$ pair, and
would have a very large width. The experimental width of $Y(2175)$
is only about $60$ MeV, which seems too narrow to be a pure
tetraquark state. We can discuss the decay of the $Y(2175)$ by
borrowing an argument based on a valence quark picture. The $(\bar s
s) (\bar s s)$ configuration for $Y(2175)$ can be a combination of
$^3S_1$ and $^3P_0$, which may fall apart into two mesons of $1^-$
and $0^+$ in the $s$-wave. In the QCD sum rule the $1^-$ $\bar s s$
meson is well identified with $\phi(1020)$, while the $0^+$ $\bar s
s$ meson has a mass around $1.5$ GeV and is hard to be identified
with the observed $f_0(980)$. Therefore, such a fall-apart decay
would simply be suppressed due to the kinematical reason. The
physical $f_0(980)$ state may be a tetraquark state as discussed in
the previous QCD sum rule study~\cite{Chen:2007zzg}. Then the
transition $Y(2175) \to \phi(1020) + f_0({\rm tetraquark})$ should
be accompanied by a $\bar q q$ creation violating the OZI rule, as
well as by an annihilation of one quanta of orbital angular
momentum. These facts may once again suppress the decay of $Y(2175)
\to \phi(1020) + f_0(980)$. This fact was studied in the recent
paper by Torres, Khemchandani, Geng, Napsuciale and
Oset~\cite{Torres:2008gy}. They studied the $\phi K \bar K$ system
with the Faddeev equations where the contained $K \bar K$ form the
$f_0(980)$ resonance. The decay width they calculated is around $18$
MeV, not far from the experimental value. The all above evidences
would imply that the $Y(2175)$ is a possible candidate of a
tetraquark state.

$Y(2175)$ could be a threshold effect, a hybrid state $s\bar s G$, a
tetraquark, an excited $s\bar s$ state or a mixture of all the above
possibilities. Because of its non-exotic quantum number, it is not
easy to establish its underlying structure. Clearly more
experimental and theoretical investigations are required.

One byproduct of the present work is the interesting observation
that some type of four-quark interpolating currents may couple
weakly to the conventional $q\bar q$ ground states. If future work
confirms this point, we may have a novel framework to study the
excited $q\bar q$ mesons using the four-quark interpolating
currents, which is not feasible for the traditional $q\bar q$
interpolating currents.

%
\section*{Acknowledgments}
%
The authors thank G. Erkol for useful discussions. H.X.C. is
grateful for Monkasho support for his stay at the Research Center
for Nuclear Physics where this work is done. X.L was supported by
the National Natural Science Foundation of China under Grants
10705001 and the China Postdoctoral Science foundation
(20060400376). A.H. is supported in part by the Grant for Scientific
Research ((C) No.19540297) from the Ministry of Education, Culture,
Science and Technology, Japan. S.L.Z. was supported by the National
Natural Science Foundation of China under Grants 10625521 and
10721063 and Ministry of Education of China.
%

%

\end{document}